\RequirePackage{amsmath}
\documentclass{svjour3}
\usepackage[T1]{fontenc}
\usepackage[latin9]{inputenc}
\usepackage{xcolor}
\usepackage{pdfcolmk}
\usepackage{url}
\usepackage{amssymb}
\usepackage{graphicx}
\PassOptionsToPackage{normalem}{ulem}
\usepackage{ulem}

\makeatletter

\usepackage{graphicx}

\makeatother

\begin{document}

\title{Can we Falsify the Consciousness-Causes-Collapse Hypothesis in Quantum
Mechanics?}

\author{J. Acacio de Barros \and Gary Oas}

\institute{J. Acacio de Barros \at School of Humanities and Liberal Studies\\
San Francisco State University\\
San Francisco, CA 94132\\
\email{barros@sfsu.edu} \and G. Oas \at Stanford Pre-Collegiate
Studies\\
Stanford University\\
Stanford, CA 94305}

\date{Received: date / Accepted: The correct dates will be entered by the
editor}
\maketitle
\begin{abstract}
In this paper we examine some proposals to disprove the hypothesis
that the interaction between mind and matter causes the collapse of
the wave function, showing that such proposals are fundamentally flawed.
We then describe a general experimental setup retaining the key features
of the ones examined, and show that even a more general case is inadequate
to disprove the mind-matter collapse hypothesis. Finally, we use our
setup provided to argue that, under some reasonable assumptions about
consciousness, such hypothesis is unfalsifiable. 
\end{abstract}

\section{Introduction}

One of the central issues within Quantum Mechanics (QM) is the measurement
problem. Though many different solutions to it have been offered (e.g.
\cite{bohm_suggested_1952,bohm_suggested_1952-1,everett_relative_1957,bohm_proposed_1966,omnes_interpretation_1994,fuchs_introducing_2014}),
there is no consensus among physicists that a satisfactory resolution
has been achieved. Perhaps the main reason for this disagreement is
the lack of clear experimental procedures that could distinguish an
interpretation from another. For example, Bohm's theory yields
exactly the same predictions as the standard Copenhagen interpretation
for quantum systems \cite{holland_quantum_1995}, at least for most
measurable quantum systems\footnote{For extreme cases where there might be some differences, albeit not
necessarily directly observable; see \cite{de_barros_causal_1998,de_barros_causal_1998-1}
or \cite{valentini_dynamical_2005}.}.

Among the proposed solutions, perhaps one of the most controversial
is von Neumann's idea that a measurement is the result of the interaction
of a (conscious) mind with matter \cite{von_neumann_mathematical_1996}.
This idea posits two distinct types of dynamics for quantum systems:
one linear, to which all matter is subject under its standard evolution,
and another non-linear and probabilistic, to which matter is subject
when it interacts with an observer's mind. This is a substance-dualist
view, where matter and mind exist in different realms and satisfy
different laws of nature. This interpretation has Henry Stapp as its
currently best-known supporter \cite{stapp_mind_2009}. We shall label 
the hypothesis that the interaction with a mind causes the collapse
of the wave function the \textit{Consciousness Causes Collapse Hypothesis}
(CCCH). 

Recently, some authors claimed that  CCCH was inconsistent with
already available empirical evidence (see, e.g. \cite{yu_quantum_2011,thaheld_can_2015}).
In this paper, we examine  CCCH with respect to such claims, in
particular those of \cite{yu_quantum_2011}, and show that their proposal
does not provide a way to falsify  CCCH. We then modify their proposal
to a stripped-down version that retains the main features of an experiment
needed to falsify  CCCH. This exposes a fundamental problem: to
test  CCCH one would need to make a conscious being part of the
experimental setup. Unless we subscribe to a panpsychist view of consciousness
(which  CCCH proponents usually do not), such types of experiment
pose a fundamental problem: to have a conscious being, one needs reasonably
high temperatures (compared to absolute zero). Thus, any experiment
that distinguishes two orthogonal states of a measurement, as we shall
see is necessary, cannot be brought to its original quantum state,
as this would imply controlling all the quantum states in a thermal
bath. Therefore, For All Practical Purposes (FAPP), the outcomes of
such experiments would be inconclusive, and they would not test 
CCCH. In fact, this suggests that, due to environmental decoherence,
 CCCH is unfalsifiable FAPP. 

We organize this paper in the following way. In Section \ref{sec:The-von-Neumann}
we briefly discuss the von Neumann interpretation of quantum mechanics.
In Section \ref{sec:Proposed-falsification-QMH} we present Yu and
Nikolic's experiment, and describe why it does not work as proposed.
Then, in Section \ref{sec:Is-QHM-falsifiable}, we modify their experimental
setup, and analyze under which conditions the modified experiment
needs to be performed to test  CCCH. We end the paper with some
conclusions. 

\section{The Consciousness-Causes-Collapse interpretation of QM\label{sec:The-von-Neumann}}

In this section, we present the idea of the consciousness-causes-collapse
interpretation, which originated from von Neuman's work on the measurement
problem in quantum mechanics. In his seminal book \cite{neumann_mathematical_1932},
von Neumann starts with the assumption that every physical system
can be represented as a vector $|\psi\rangle$ in a Hilbert space
$\mathcal{H}$. This representation is one-to-one, in the sense that
not only every system has a corresponding vector, but that to every
vector there is, in principle, a corresponding system. Observable
quantities are represented in this Hilbert space as linear Hermitian
operators. The spectral decomposition theorem tells us that a Hermitian
operator $\hat{A}$ can be written as 
\[
\hat{A}=\sum_{i}a_{i}\hat{P}_{i},
\]
where $a_{i}\in\mathbb{R}$ and $\hat{P}_{i}$ are projection operators
such that $\hat{P}_{i}\hat{P}_{j}=\delta_{ij}\hat{P}_{j}$. In von
Neumann's view, the dynamics of a system is more complicated, and
we should distinguish two types. One type is given when the system
does not interact with a measurement device. When this is the case,
the evolution of the state $|\psi\rangle$ follows a deterministic
and linear evolution given by Schr\"odinger's equation. Namely, the
state of the system at time $t_{1}\geq t_{0}$ is given by 
\[
|\psi\left(t_{1}\right)\rangle=\hat{U}\left(t_{1};t_{0}\right)|\psi\left(t_{0}\right)\rangle,
\]
where $\hat{U}\left(t_{1};t_{0}\right)$ is a unitary evolution operator
between $t_{0}$ and $t_{1}$ given by 
\[
\hat{U}\left(t_{1};t_{0}\right)=\exp\left[-\frac{i}{\hbar}\hat{H}\left(t_{1}-t_{0}\right)\right],
\]
and $\hat{H}$ is the Hamiltonian operator. If, on the other hand,
the system interacts with a measurement device, the evolution is not
linear nor deterministic. During a measurement, each observable value
$a_{i}$ has a probability $p\left(a_{i}\right)=\left|\hat{P}_{i}|\psi\rangle\right|^{2}$
of being observed, and if the result of a measurement (with probability
$p\left(a_{i}\right)$) is $a_{i}$, then the wave-function collapses
into a new state
\[
|\psi\rangle\xrightarrow{a_{i}}\frac{\hat{P}_{i}|\psi\rangle}{\langle\psi|\hat{P}_{i}|\psi\rangle}.
\]
So, according to this formulation, QM has two different types of evolution,
one deterministic and one probabilistic; the former happens when there
is no interaction with a measurement device, and the latter when such
interaction occurs. 

A natural question to ask within this theory is ``what is a measurement
device?'' In principle, such a device, made out of ``conventional''
matter itself, should be describable by QM. Following von Neumann,
let us assume this is the case, and let us have a Hilbert space 
$\mathcal{H}=\mathcal{H}_{M}\otimes\mathcal{H}_{S}$,
where $\mathcal{H}_{M}$ is the space of the measurement device and
$\mathcal{H}_{S}$ the space of the system being measured. Since we
are considering this an isolated system, there is no interaction with
an external measuring device (the device is part of the system itself).
For simplicity, let us limit our measuring device to the observable
\[
\hat{O}=\hat{P}-\left(\hat{1}-\hat{P}\right)=2\hat{P}-\hat{1},
\]
where $\hat{P}^{2}=\hat{P}\neq\hat{1}$ is a projector, and $\hat{1}$
the identity operator. Clearly, $\hat{O}$ can have only two possible
outcomes, $+1$ and $-1$. So, a measuring device for $\hat{O}$ needs
to have the following properties. First, it should have a neutral  
state, its initial state, prepared to receive a system to be measured.
We denote the neutral state of the measuring device by the vector
$|\mbox{neutral}\rangle\in\mathcal{H}_{M}$. Second, the interaction
of $M$ and $S$ should be such that the following evolution happens:
\[
|\mbox{neutral}\rangle\otimes|+\rangle
\rightarrow
\hat{U}_{\mbox{int}}|\mbox{neutral}\rangle\otimes|+\rangle = |\mbox{points to }+\rangle\otimes|+\rangle,
\]
\[
|\mbox{neutral}\rangle\otimes|-\rangle
\rightarrow
\hat{U}_{\mbox{int}}|\mbox{neutral}\rangle\otimes|-\rangle = |\mbox{points to }-\rangle\otimes|-\rangle.
\] 
Here we represent the two possible final values of the measurement
apparatus as either giving a measurement of ``$+$'' or ``$-$,''
depending on the initial state of the system. 

Since, according to QM, any linear superposition of states $|\pm\rangle\in\mathcal{H}_{S}$
is possible, what happens when we use the above interaction to measure
superpositions? If we have the superposition
\[
|\psi\rangle=c_{+}|+\rangle+c_{-}|-\rangle,
\]
because $\hat{U}_{\mbox{int}}$ is linear, it follows that 
\[
|\mbox{neutral}\rangle\otimes|\psi\rangle\rightarrow 
c_{+}|\mbox{points to }+\rangle\otimes|+\rangle+c_{-}|\mbox{points to }
-\rangle\otimes|-\rangle.
\]
This seems to be exactly what we wanted: we end up with a correlation
between $|\pm\rangle$ and the pointer's state $|\mbox{points to }\pm\rangle$.
However, it is straightforward to see that the final state is \emph{not}
an eigenstate of either projector $\hat{1}\otimes|+\rangle\langle+|$
or $\hat{1}\otimes|-\rangle\langle-|$, and therefore does not correspond
to an actual measurement, where an actual collapse happens. This contains
the essence of the measurement problem: a quantum system interacting with
a measurement apparatus evolves according to a non-linear dynamics
that is different from that given by the (linear) Schroedinger equation. 

If the quantum system was in a superposition,  von Neumann 
argued that  the interaction of $S$ with a measurement apparatus
$M$ would also result in a superposition. We could push this even 
further and think of another
apparatus $M'$ that measures $M$ and $S$, and we would still have a
superposition. In fact, we could keep doing this indefinitely, ever adding more
measurement apparatuses that measure the previous measurement devices. 
We can even consider our eyes as a photodetector
that measures this chain of apparatuses, and we have no
reason to assume, according to Schroedinger's equation, that we would not have
a superposition. We can keep on going, including not only our eyes,
but our optical nerves, up until we get to the brain, and we are left
with a brain/measurement apparatus/system that is still in a superposition. 
In von Neumann's own words:
\begin{quote}
``That this boundary {[}between observer and observed system{]} can
be pushed arbitrarily deeply into the interior of the body of the
actual observer is the content of the principle of the psycho-physical
parallelism \textemdash{} but this does not change the fact that in
each method of description the boundary must be put somewhere, if
the method is not to proceed vacuously, i.e., if a comparison with
experiment is to be possible. Indeed experience only makes statement
of this type: an observer has made a certain (subjective) observation;
and never any like this: a physical quantity has a certain value.''
\end{quote}

That is intriguing, and since one never observes a superposition 
in a single measurement, this chain needs to stop somewhere.

Following the consequences of von Neumann's ideas, London and Bauer
pushed the boundary to the extreme (the reader is referred to the
excellent historical survey provided in \cite{freire_junior_quantum_2015}).
According to them, there is only one step when we know \emph{for sure}
that we do not have a superposition: when we gain conscious knowledge
of the measurement apparatus, i.e. when matter interacts with the
mind . That is because we are never aware of observing any quantum
superposition. They then proposed that the interaction between mind
and matter causes matter to evolve probabilistically, according to
Born's rule, and non-linearly. In other words, the mind causes the  
collapse of the wave function. 

CCCH is substance dualist. As is well-known, dualist views  
of the mind suffer the problem of causal closure: how can the mind
influence matter and \emph{vice versa}? Though not directly addressing
this issue,  CCCH states that the mind causes matter to behave
differently, following a dynamics that is not the same as when there is no interaction
with a mind. So, in a certain sense,  CCCH postulates their
interaction, albeit in a very specific way. The question remains as
to whether this interaction may be used to actually provide a way
for the mind to affect matter in a (consciously) controlled way. 

Henry Stapp proposed a clever solution to this problem by using the
``inverse'' Quantum Zeno Effect \cite{stapp_mind_2014}. It would go beyond the scope of this paper
to provide a detailed account of Stapp's theory, but it is worth mentioning it
to give an idea of what types of physics (or metaphysics) may unfold
from the CCCH. In Ref. \cite{misra_zenos_2008}, it was shown that if we were to continuously
observe an unstable particle, this particle would not decay; this
came to be known as the Quantum Zeno Effect (QZE). The QZE can be modified,
and it can be shown that by continuous and variable observations it
is possible to \emph{force} a particle to change its quantum state.
Following this idea, Stapp \cite{stapp_mind_2014} used a 
harmonic oscillator in a coherent  
state\footnote{The coherent state $|\alpha\rangle$ of a harmonic oscillator behaves,
in some sense, in a similar way to its classical counterpart. For
instance, its expected value also oscillates with the same frequency
as a classical oscillator, and with amplitude of oscillation $\alpha$.
Coherent states are of great importance in quantum optics; see e.g.
\cite{walls_quantum_1994}.} with amplitude $\alpha$, given by the ket 
\[
|\alpha\rangle=e^{-\left|\alpha\right|^{2}/2}\sum\frac{\alpha^{n}}{\sqrt{n!}}|n\rangle,
\]
where $|n\rangle$ is an eigenvector of the number operator $\hat{N}=a^{\dagger}a$
with eigenvalue $n$, and showed that if we start in this state \emph{and
} if our mind chooses to observe it, we end with a new amplitude $\beta>\alpha$,
whereas if it chooses not to observe, the state maintains amplitude
$\alpha$. In other words, the effect of the mind ``observing'' a
system can make it change its state from $|\alpha\rangle$ to $|\beta\rangle$,
$\beta>\alpha$. There might be some (surmountable) problems with
this model, discussed in more detail in \cite{de_barros_quantum_2015,de_barros_model_2016},
but we emphasize that the CCCH, though not popular among physicists and presenting
some difficult philosophical challenges, not only solves the measurement
problem, but also provides a possible mechanism for the mind to affect
matter, a major problem for substance dualists. 

\section{A proposed falsification of the CCCH\label{sec:Proposed-falsification-QMH}}

It is reasonable
to ask whether CCCH is true or false. By true or false we of course
mean whether there is supporting experimental evidence for it or if
it can be or has been falsified, as we cannot, in a strict sense,
prove a theory to be true. So, an natural question is how can
we try to falsify  CCCH. 

In a recent paper \cite{yu_quantum_2011}, Yu and Nikolic argued that
 CCCH has already been falsified, and proposed further modifications
of a given experimental setup to make such conclusions beyond any
reasonable doubt. Their argument starts with the idea that 
\[
\text{CCCH}\rightarrow\left(\text{CWF}\iff\text{PR}\right),
\]
where CWF is short for ``collapse of the wave function'' and PR
for ``phenomenal representation,'' i.e. the presence of phenomenal
consciousness. Therefore, they conclude, if it is possible to ``observe''
CWF without PR, then CCCH is falsified. 

\begin{figure}
\begin{centering}
(a)\includegraphics[scale=0.7]{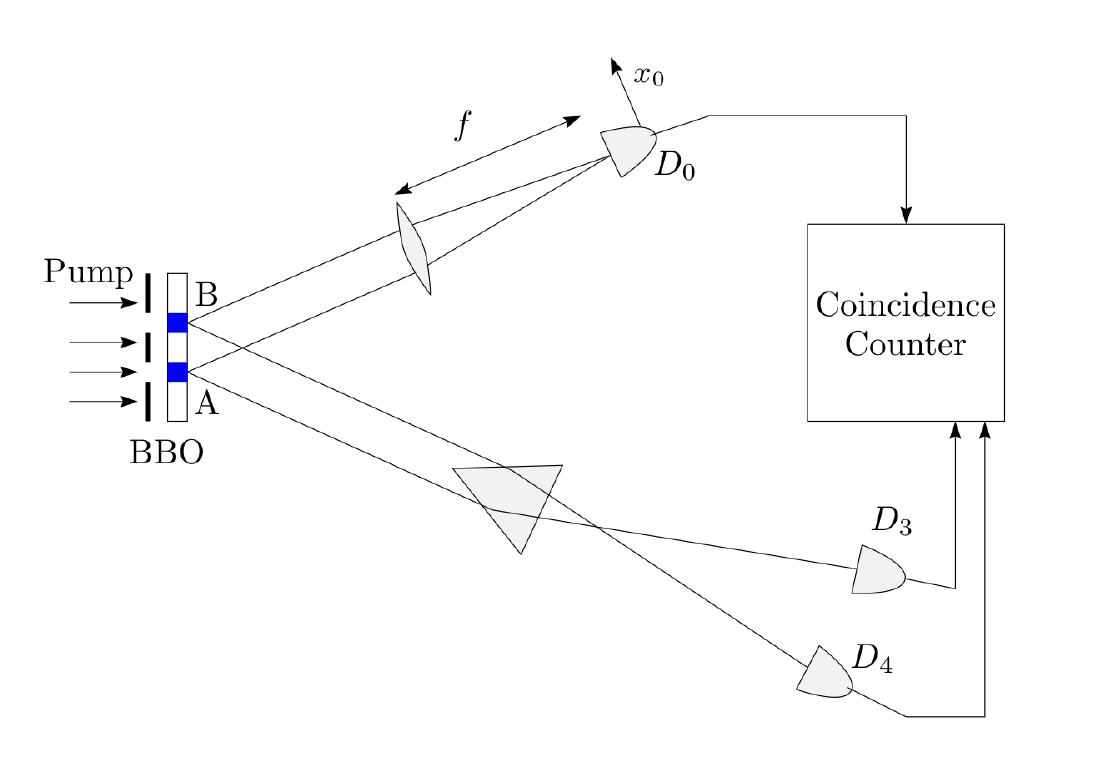}
\par\end{centering}
\begin{centering}
(b)\includegraphics[scale=0.7]{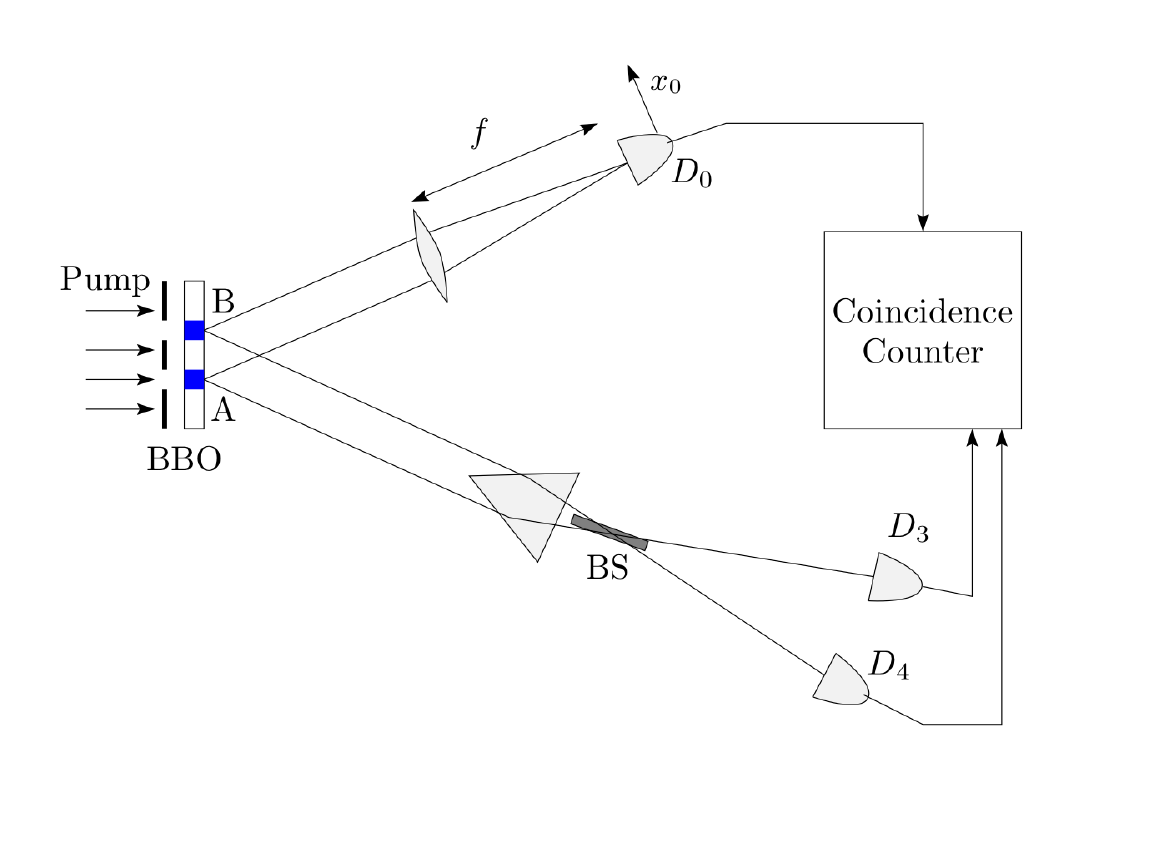}
\par\end{centering}
\caption{\label{fig:Kim-et-al-figure}Kim et al. experimental setup for the
delayed choice quantum eraser \cite{kim_delayed_2000}. }
\end{figure}

To understand Yu and Nikolic's argument, and our criticism of it,
we need to look into the details of how they account for the
possibility of observing CWF without PR. They do so by using Kim et
al.'s delayed choice experiment \cite{kim_delayed_2000}, which we
now describe. In Kim et al. (see  Figure \ref{fig:Kim-et-al-figure}), 
a  laser beam impinges on a standard
double slit, behind which a non-linear  crystal is placed. Through
parametric down conversion, a pair of photons,  referred to as 
signal and idler, is generated in either
region $A$ or $B$ of the crystal residing behind each slit.
The signal photon is sent to a detector $D_{0}$ that can be translated to 
reveal an interference pattern. The idler photon 
can be directed directly to either detector 
$D_{3}$ or $D_{4}$ (Figure \ref{fig:Kim-et-al-figure} (a)), thus
allowing which-path information, or can be scrambled in a beam splitter
$BS$ (Figure \ref{fig:Kim-et-al-figure} (b)), erasing any which-path
information. 

\begin{figure}
\begin{centering}
\includegraphics[scale=0.15]{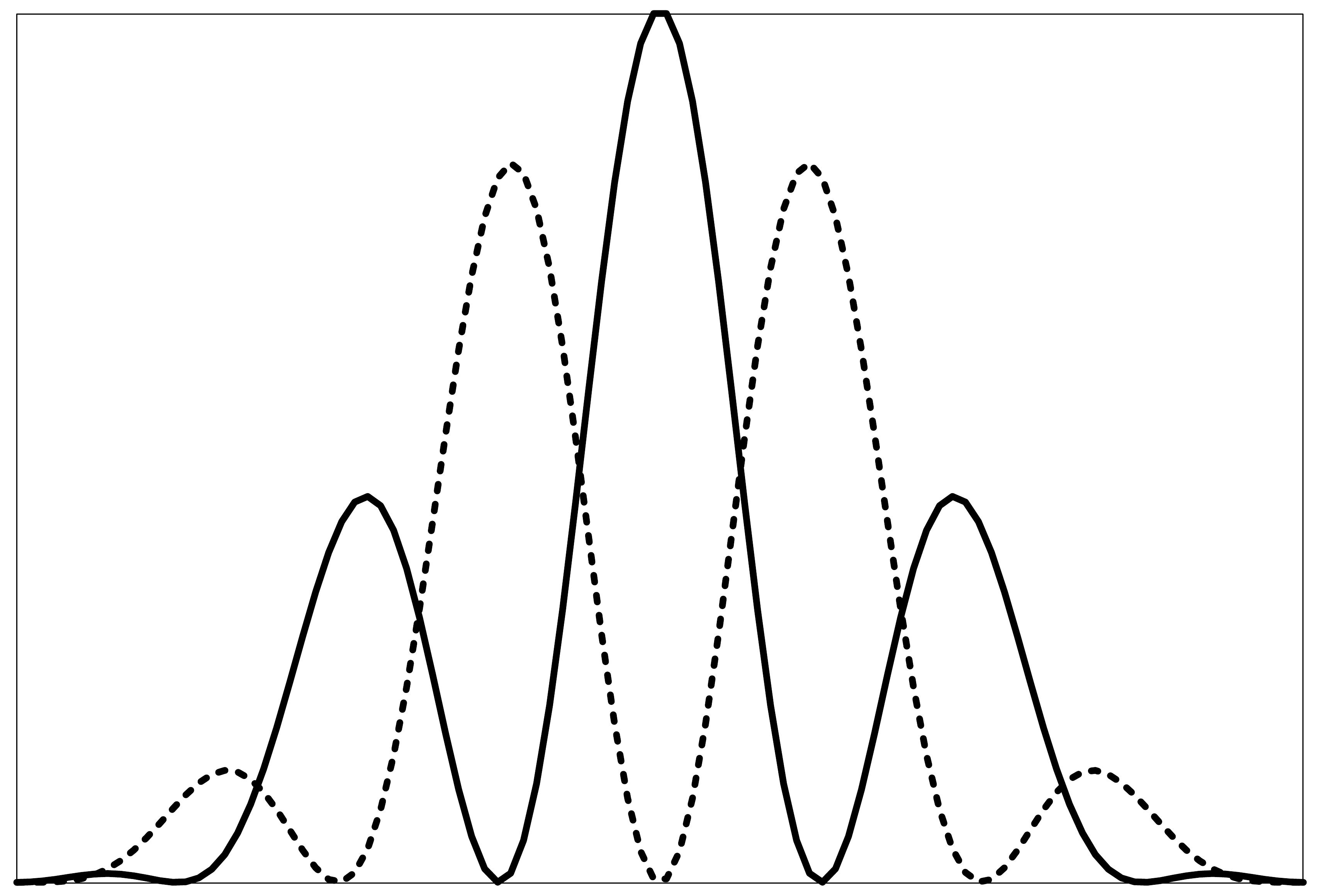}
\par\end{centering}
\caption{\label{fig:Conditional-probabilities}Probabilities $P\left(D_{0},x|D_{i}\right)$
of observing a photon in detector $D_{0}$ positioned at $x$, conditioned
on a detection on $D_{3}$ (solid line) or $D_{4}$ (dashed line). }
\end{figure}

\begin{figure}
\begin{centering}
\includegraphics[scale=0.15]{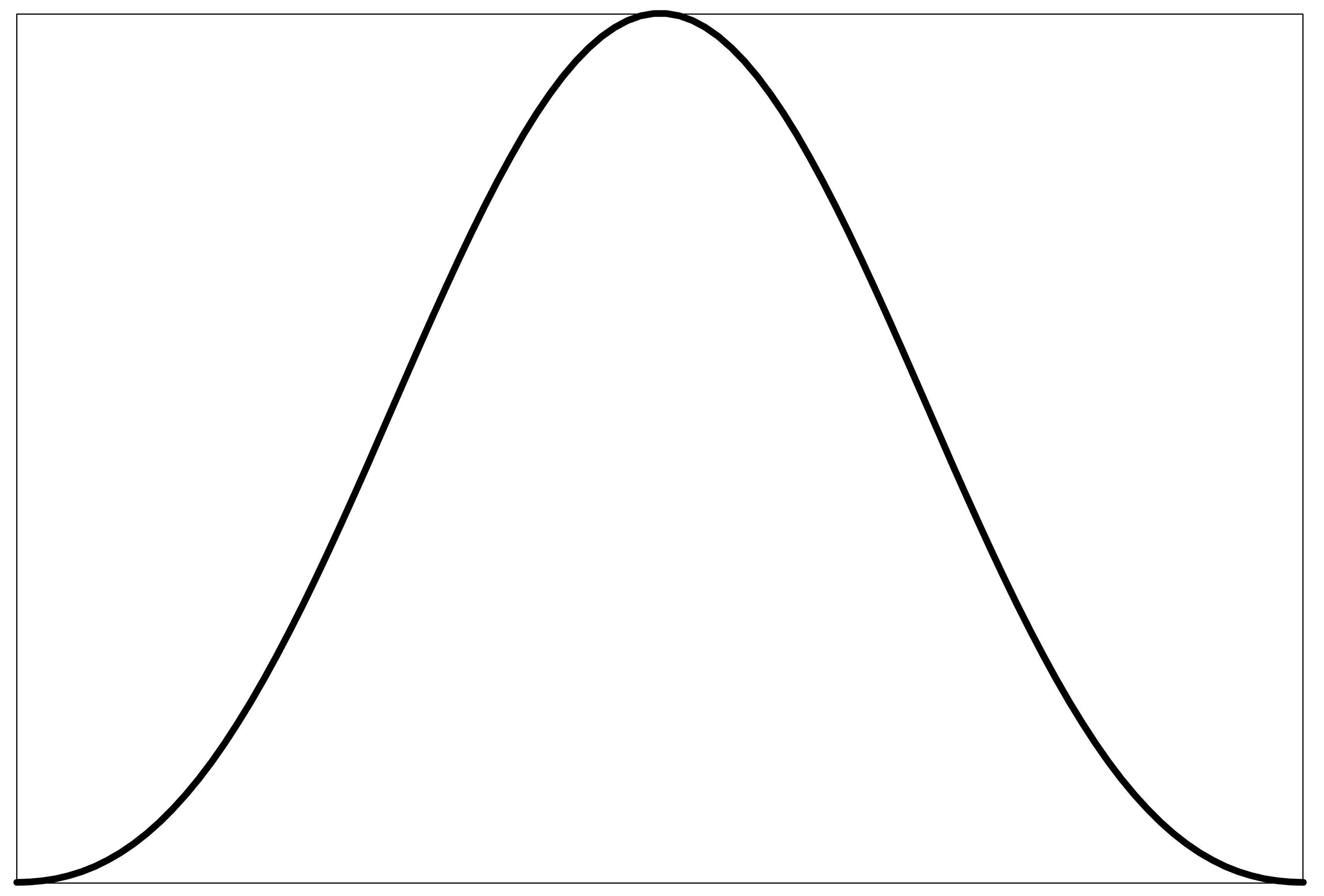}
\par\end{centering}
\caption{\label{fig:Unconditional-probabilities}Probability as a function
of $x$ of observing a photon in detector $D_{0}$ positioned at $x$. }
\end{figure}

To understand Kim et al.'s experiment, it is important to notice first
that it is a fourth-order interference experiment\footnote{Readers 
not familiar with fourth-order interference are encouraged
to consult \cite{ou_quantum_1988} or one of the many excellent textbooks
on quantum optics, such as \cite{mandel_optical_1995}.}. Let us 
analyze what happens in each of the setups (for details relevant
to the experiment discussed here, see, e.g. \cite{rubin_theory_1994}).
First, for the which-path information setup in Figure \ref{fig:Kim-et-al-figure}
(a), there is nothing unusual. The pair of photons is produced either
in $A$ or $B$, and if it is produced in $A$ the idler photon is
detected in $D_{3}$, and if in $B$ it is detected in $D_{4}$. Since
the signal photon is generated in either $A$ or $B$, the final probability
of observing it in the variable-position detector $D_{0}$ is the
same as the sum of the two probabilities, and shows no interference
effect, as expected. For the interference setup shown in Figure \ref{fig:Kim-et-al-figure}
(b), things are more subtle, and the experimental setup resembles,
conceptually, what happens with ghost interference (another fourth-order
interference experiment) \cite{strekalov_observation_1995}. When
the idler photons from $A$ or $B$ are joined, we lose which path
information, but, more importantly, the idler side of the apparatus
becomes an interference device itself, sensitive to the momentum of the
quantum state impinging on it. Different momenta, which are correlated
with $D_{0}$, produce different interference patterns in $D_{0}$,
and the overall probability distribution observed in $D_{0}$ is exactly
the same as with setup (a). As a consequence, the \emph{conditional}
probability of detection on $D_{0}$ depends on a detection on $D_{3}$
or $D_{4}$ in the following way \cite{kim_delayed_2000}:
\begin{equation}
P\left(D_{0},x|D_{3}\right)=N\left(\alpha x\right)^{-2}\sin^{2}\left(\alpha x\right)\cos^{2}\left(\beta x\right),\label{eq:d0conditiond1}
\end{equation}
\begin{equation}
P\left(D_{0},x|D_{4}\right)=N\left(\alpha x\right)^{-2}\sin^{2}\left(\alpha x\right)\sin^{2}\left(\beta x\right),\label{eq:d0conditiond2}
\end{equation}
where $N$ is a normalization factor, and $\alpha$ and $\beta$ parameters
that depend on the optical geometry of the experiment and the correlated
photons wavelength. The two conditional probabilities in (\ref{eq:d0conditiond1})
and (\ref{eq:d0conditiond2}) are shown in Figure \ref{fig:Conditional-probabilities}.
 As we can see, by \emph{conditioning} the data on the detection
of, say, $D_{3}$, we observe an interference pattern, and likewise
for the conditioned data on $D_{4}$. However, as we can also
see from Figure \ref{fig:Conditional-probabilities}, the interference
pattern obtained by conditioning on $D_{3}$ is shifted by $\pi/2$  with respect
to the one from $D_{4}$ (this is also clear from (\ref{eq:d0conditiond1})
and (\ref{eq:d0conditiond2})). This is a crucial point: 
the interference pattern does not appear on $D_{0}$
without correlating it with the detections on $D_{3}$ or $D_{4}$.
In fact, if we look only at $D_{0}$, what we see is the unconditional
$P\left(D_{0},x\right)$, given by $P\left(D_{0},x|D_{3}\right)P\left(D_{3}\right)+P\left(D_{0},x|D_{4}\right)P\left(D_{4}\right)$,
shown in Figure \ref{fig:Unconditional-probabilities}. 
 If this were not the case, we would violate the no-signaling condition
in quantum mechanics, as we could use a choice of detection apparatus
in $D_{i}$ to communicate instantaneously (or to the past) between
an experimenter controlling $D_{i}$ and another observing $D_{0}$.
But since the observations are conditional, no violation of no-signaling
occurs. 

Returning to Yu and Nikolic's idea, their proposal was to use the
human eye as a photodetector instead of $D_{i}$. This would not be
an impossible task, given that human eyes are sensitive to single
photons. As such, they argue that, in the which-path setup where the
idler photon goes to $D_{3}$ and $D_{4}$, we could replace detectors
$D_{3}$ and $D_{4}$ with a person observing the photons. If such
observer were unconscious, then no collapse of the wave function would
happen, and we would have an interference pattern on $D_{0}$.  Notice that 
Yu and Nikolic are referring to setup (a) in  Figure \ref{fig:Kim-et-al-figure},
and they do not consider setup (b), where  interference patterns 
emerge in Kim et al.'s experiment. As such, their proposal has 
a major flaw: one would not get an interference pattern on $D_{0}$ using
setup (a) regardless of having a detector or an observer
(conscious or not). To obtain 
an interference pattern, any which-path information  
about the idler photon needs not only to be erased by recombining
the beams into an interferometer, but once recombined one would need
to detect such photon and use coincidence counts to obtain the interference.
If one used an actual person to observe $D_{3}$ or $D_{4}$, such
coincidence counts could only happen if such person was aware of the
detection in their eye, as this would be required for knowing which
detections in $D_{0}$ need to be counted. In other words, a human
(or any other animal) used in this experimental setup would have to
be aware of the detection of a photon within a certain window of time
and be able to behaviorally track (e.g. by recording on a piece of paper) such detection,
such that later on an interference pattern could be reconstructed
by coincidence counts\footnote{In fact, the 
total number of photons reaching the participant (either
human or not) is quite large, and it is not until coincidence counts
are performed that this number is reduced. So, the task of reconstructing
an interference pattern, even if the actual photon count per second
could be reduced to a reasonable number to be dealt with, would be
very time consuming and daunting. }. 

An interesting question is raised from Yu and Nikolic's proposal:
could we falsify  CCCH with some device of this type? As we saw,
their claim that  CCCH was (perhaps) already falsified is not correct,
as their reliance on the quantum eraser experiment did not take into
consideration the need for correlated counts. But perhaps some other
version of the experiment could to it. In the next section we will
show a general type of experiment to test  CCCH, and use it to
argue that it is impossible to falsify  CCCH. 

\section{Is  CCCH falsifiable?\label{sec:Is-QHM-falsifiable}}

In this section we describe a different proposed experiment to test
 CCCH. This experiment is a natural extension of an earlier paper
of Suppes \& de Barros \cite{suppes_quantum_2007}, and has the main
features necessary to test  CCCH. Our goal here is not to propose
a thought experiment, but to examine the characteristics of a realizable
experiment, and discuss its conceptual and technical difficulties. 

Since we want to test  CCCH, like Yu and Nikolic, we start with
the eyes as photo-detectors. Nature provides us with exceptionally
good photo-detectors in the kingdom of \emph{Animalia} (see references
in \cite{suppes_quantum_2007}). Of particular interest, is the fact
that some insects have not only very efficient eyes (their efficiency
is estimated to be between 40\% and 78\%), but very low dark-count
rates (the locust \emph{Schitocerca gregaria}, for example, has a
dark-count rate of few photons per hour). 

Perhaps one of the best candidates for such conditioning experiments
is the cockroach (\emph{Periplaneta americana}), for the following
reasons \cite{lent_antennal_2004}: it responds well to external stimuli
for conditioning, it is well adapted to respond to very low-light
environments (i.e. has good photo-detectors), and its neural circuitry
is significantly easier to study compared to other well-known insects
(such as the ubiquitous fruit fly). So, for that reason, in combination
with the existence of successful conditioning experiments with insects,
Suppes \& de Barros \cite{suppes_quantum_2007} proposed that cockroaches
could be classically conditioned to respond to single photons. 

Here we assume that cockroach single-photon conditioning experiments
could be successfully carried out, though probably there exists many
technical difficulties (insect are not as easy to condition as some
mammals). For our purpose, we will also assume that the cockroach
is a conscious being. This is, of course, a controversial assumption,
but the alternative would be to do our proposed experiment with more
complex animals (say, humans). However, as it will become clear below,
this assumption will not invalidate our conclusions, as they will
apply to any animal. 

The idealized experiment we propose is simple, and does not rely on
entangled states (as does Kim et al.'s). Imagine we have a cockroach
who has been conditioned to respond to single photons in the following
way. If a photon impinges on the left eye of the cockroach, it moves
its left antenna, whereas if a photon impinges on the right eye it
moves its right antenna. 
\begin{figure}
\centering{}
\includegraphics[viewport=0bp 200bp 612bp 650bp,clip,scale=0.3]{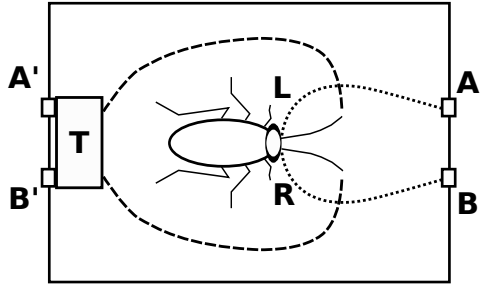}
\caption{Proposed experimental setup. A photon 
impinges on A or B, and an optical
fiber, represented by the dotted line, takes it to either the left
(L) or right (R) eye, respectively. If a photon reaches L, the cockroach
is conditioned to push a button at the end of a circuit (dashed line),
and if the L button is pushed, a single photon is emitted at a precise
and very short window of time. }
\end{figure}
 The cockroach is then placed in a well isolated box where a photon
can be sent to either the left or the right eye via optical fibers.
If the cockroach's left antenna moves, the cockroach sends a signal
to a device T that will generate a single photon from A'; if the right
antenna moves, a single photon is generated from B'. Now, the idea
here is that if instead of a single photon in A or B, a quantum superposition
$|\psi\rangle=c_{1}|1\rangle_{A}|0\rangle_{B}+c_{2}|0\rangle_{A}|1\rangle_{B}$
was sent to the box, the output would be a quantum superposition if
the cockroach is not conscious, whereas it would be a proper mixture
if the cockroach caused a collapse of the wave function. 

Now, to understand the experimental conditions necessary for such
experiment to work, let us examine it in detail. We start with the
Hilbert space of this setup, given by $\mathcal{H}=H_{p}\otimes H_{c}\otimes H_{b}\otimes H_{p'}$,
where $H_{p}$ is the Hilbert space for the impinging photon, $H_{c}$
the cockroach, $H_{b}$ the box itself (with all necessary devices),
and $H_{p'}$ the outgoing photon. For example, when a single photon
impinges on A, with 
\[
\rho_{1,0}=|1_{A},0_{B}\rangle\langle1_{A},0_{B}|,
\]
the initial state of the system is given by
\[
\rho_{1,0}\otimes\rho_{\text{ready}}^{\text{roach}}\otimes
\rho_{\text{ready}}^{\text{box}}\otimes\rho'_{0,0},
\]
where 
\[
\rho_{\text{ready}}^{\text{roach}}=|\text{cockroach 
ready}\rangle\langle\text{cockroach ready}|,
\]
\[
\rho_{\text{ready}}^{\text{box}}=|\text{box ready}\rangle\langle\text{box ready}|,
\]
and 
\[
\rho'_{0,0}=|0_{A'},0_{B'}\rangle\langle0_{A'},0_{B'}|.
\]
This system would evolve the following way:
\begin{align*}
\rho_{1,0}\otimes\rho_{\text{ready}}^{\text{roach}}\otimes
\rho_{\text{ready}}^{\text{box}}\otimes\rho'_{0,0} & \rightarrow\\
\rho_{0,0}\otimes\rho_{\text{left antennae}}^{\text{roach}}\otimes\rho_{\text{ready}}^{\text{box}}\otimes\rho'_{0,0} & \rightarrow\\
\rho_{0,0}\otimes\rho_{\text{ready}}^{\text{roach}}\otimes\rho_{\text{gen.photon A'}}^{\text{box}}\otimes\rho'_{0,0} & \rightarrow\\
\rho_{0,0}\otimes\rho_{\text{ready}}^{\text{roach}}\otimes\rho_{\text{ready}}^{\text{box}}\otimes\rho'_{1,0},
\end{align*}
where the label for the states should make them evident. A similar
evolution would happen to $\rho_{0,1}$, leading to 
\begin{align*}
\rho_{0,1}\otimes\rho_{\text{ready}}^{\text{roach}}\otimes\rho_{\text{ready}}^{\text{box}}\otimes\rho'_{0,0} & \rightarrow\\
\rho_{0,0}\otimes\rho_{\text{ready}}^{\text{roach}}\otimes\rho_{\text{ready}}^{\text{box}}\otimes\rho'_{0,1}.
\end{align*}
Finally, if we started with a superposition given by, say, the state
\[
\rho_{1,1}=\frac{1}{2}\left(|0_{A},1_{B}\rangle\langle0_{A},1_{B}|+|1_{A},0_{B}\rangle\langle0_{A},1_{B}|+|0_{A},1_{B}\rangle\langle1_{A},0_{B}|+|1_{A},0_{B}\rangle\langle1_{A},0_{B}|\right),
\]
we would end with the linear evolution 
\begin{align*}
\rho_{1,1}\otimes\rho_{\text{ready}}^{\text{roach}}\otimes\rho_{\text{ready}}^{\text{box}}\otimes\rho'_{0,0} & \rightarrow\\
\rho_{0,0}\otimes\rho_{\text{ready}}^{\text{roach}}\otimes\rho_{\text{ready}}^{\text{box}}\otimes\rho'_{1,1}.
\end{align*}
Clearly, if the experiment could be performed like above, if the input
is a superposition, we can take the partial trace over all other variables,
and the output will also be a superposition. In other words, because
the evolution is linear, the partial trace over $H_{p}\otimes H_{c}\otimes H_{b}$
of $\rho_{0,0}\otimes\rho_{\text{ready}}^{\text{roach}}\otimes\rho_{\text{ready}}^{\text{box}}\otimes\rho'_{1,1}$
would result in $\rho'_{1,1}\in H_{p'}$. However, if the cockroach's
mind causes a collapse of the wave function inside the box, then the
dynamics would not be linear, and the output would be the proper mixture
\[
\rho_{\text{mixture}}^{'}=\frac{1}{2}\left(|1_{A'},0_{B'}\rangle\langle1_{A'},0_{B'}|+|0_{A'},1_{B'}\rangle\langle0_{A'},1_{B'}|\right),
\]
 and not the pure state $\rho'_{1,1}$. 

However, from the system's evolution above, we can see a major difficulty
with such an experiment, which also will plague any other experiment
attempting to falsify the CCCH. In order for a superposition to be
detected at the output, the cockroach and box need to go back to its
original \emph{quantum} state. It is easy to see, for instance, that
if the cockroach does not go back to its original state $\rho_{\text{ready}}^{\text{roach}}$,
then the final state would be an entanglement between the different
cockroach positions for inputs A or B. Then, if the outside experimenter
observes this system (causing its collapse?), what they would see
is a proper mixture, and not a superposition. Therefore, for such
an experiment to work in testing  CCCH, the whole cockroach+box
needs to be brought back to its original state\footnote{To be more 
precise, elements in the Hilbert space that are not entangled
with the original photon state need not return to the original quantum
state. Furthermore, for elements that are weakly entangled it may
not be necessary to return them to the original state either, though
not returning them would reduce the visibility of the quantum superposition.
However, this is not essential for the arguments that follow, since
the number of degrees of freedom that get entangled correspond to
a macroscopic portion of the cockroach. }. This means that every 
single atom that makes up the cockroach, for
example, needs to be brought back to its original state. Of course,
though a tremendously difficult task, it is not forbidden by quantum
mechanics (though, not  experimentally feasible, FAPP). 

An attentive reader  may counter-argue that  a carefully designed 
experiment, where all degrees of freedom are followed, would allow for 
the differentiation between CCCH and its negation (if we accept the 
assumption that the cockroach is conscious).  For example,  one would 
not need to partial trace over the system to obtain the photon outcome 
in a superposition state: we could  simply
observe the whole system (cockroach + photon + box), and see that, 
if  CCCH is false, it would be in a quantum superposition.  For instance, 
this is similar to what is done in some 
recent Schr\"odinger "kitten" experiments,  where mesoscopic 
systems are placed in a superposition state \cite{bruno_displacement_2013}.  
In fact, some researchers even proposed to create superposition states of  
bacteria  \cite{li_quantum_2016} and even macroscopic 
living organisms, such as the {\em tardigrade} \cite{Zhang-qi_bringing_2017}.
However, we must point out that all of those proposals have in common a very 
weak (and controlled) coupling with the environment, and usually
at very low temperatures. For example, what makes the {\em tardigrade}
interesting for this type of experiment is that it is able to survive
in a  {\em vacuum} for short periods of time as well as  very low temperatures, 
close to absolute zero.  Such low temperatures are necessary to decrease
the coupling of the {\em tardigrade} with the thermal environment, and 
one may even argue that while in a superposition the {\em tardigrade} is not
clearly "alive," less even "conscious," but certainly unable to provide a behavioral 
response, a requisite of any experimental setup similar to the one
provided above.  It is also important to note that for the experiments 
with Schr\"odinger kittens there is no measurement of the 
entire system, as the thermal environment is not measured. 

We could try to circumvent the difficulty of thermal 
coupling of a large macroscopic system by focusing only on
elements of the cockroach that are directly involved with the stimulus
and response process. For instance, if we include only the perceptual
and response systems of the cockroach, the number of particles that
would need to be controlled and brought back to the original state
is smaller than the totality of the cockroach. But if we do that,
we should expect about $10^{20}$ atoms (not including the numerous 
photons) to be involved in such process,
and the relevant subspace of the Hilbert space would still be extremely
large.  As mentioned in the previous paragraph, in order to perform 
such types of experiment with reasonable candidates
for having phenomenal representation (a cockroach is already somewhat
a questionable one), we need to decouple this system from the thermal
bath. This is a necessary strategy to create quantum superpositions, 
as in this case, of living systems: their temperature needs to be
lowered to a few kelvin. 

It is questionable whether cockroaches or {\em tardigrades} are conscious,
but any candidate for phenomenal consciousness\footnote{Unless 
we take a panpsychist view, which would, in the case of CCCH
raise other problems.} is a living creature, and as such they cannot 
have consciousness, much less  can move, at temperatures close
to absolute zero, as required for quantum superposition experiments.  
Therefore, if we include the thermal bath on the description of the
system above, even if we could bring the cockroach+apparatus back
to its original quantum state, the outcome of the experiment would
be irreversibly entangled with the thermal bath, and we would always
observe at the end a proper mixture, \emph{regardless }of whether
the cockroach caused a collapse or not. Since a thermal bath is a
necessary condition for a living candidate to have phenomenal consciousness,
 CCCH is unfalsifiable.

\section{Conclusions\label{sec:Conclusions}}

 CCCH is arguably one of the most controversial solutions for the
measurement problem in quantum mechanics, and it certainly does not
share wide support within the foundations of physics community  
\cite{schlosshauer_snapshot_2013}.  
We understand here the \emph{measurement problem} as the need to explain
how the transition between the quantum description of a physical system
and the classical description of the measuring apparatus comes to
be, since such transition does not come from the dynamics of quantum
theory (i.e., the unitary evolution given by Schr\"odinger's equation).
In such sense,  CCCH achieves this goal, albeit in a way that is
unappealing to most physicists, because of its substance-dualistic
nature. This raises deep philosophical problems, as it brings extra
metaphysical entities into play. However, this problem is not exclusive
to CCCH. For example, the many-worlds interpretation of QM postulates
the existence of an infinite number of parallel universes. Bohm's
theory, another popular interpretation, requires a physical reality
that unfolds in an infinite-dimensional universe, and provides no
clear explanation as to why we perceive a three-dimensional universe.
In fact, all well-known interpretations bring extra metaphysical entities
into play, with the exception perhaps of epistemic interpretations,
who  avoid such types of discussion. 

Given its metaphysical implications, it is not surprising that CCCH 
is often criticized, but mostly on metaphysical grounds (as
are many of the different interpretations of QM). However, 
if the mind plays a special role in the measurement process,
perhaps we can use this to create experiments where one could try
to falsify CCCH. In this paper we examined one experiment 
proposed by Yu and Nicolic \cite{yu_quantum_2011}. We saw that
their proposal had a fatal flaw, as it did not consider the fact that
to observe fourth-order interference requires coincidence counting.
We then used this experiment as a springboard to a more general
framework for how to attempt to falsify CCCH: produce an experimental
setup where the non-linear nature of the quantum dynamics in the presence
of consciousness can be distinguished from the linear dynamics in
the absence of consciousness. 

Another argument put forth against the CCCH was given by Thaheld \cite{thaheld_can_2015},
where the Stark-Einstein law was used to argue that classical  
information is passed to the eye-brain system via absorption of photons
by the retinal molecules. We will not go into the details of Thaheld's
argument, since they are not required here, but we want to point
out that the classical information is passed because of an entanglement
between a photon and the ``classical'' environmental variables,
and also that the Stark-Einstein law assumes, deep down, a collapse
of the wave function (either photon is absorbed by the molecule or
not). Thaheld's argument against the CCCH also suffers from the same
issues as the proposal put forth in Section \ref{sec:Is-QHM-falsifiable}. 

Finally, we emphasize that any candidate for phenomenal 
consciousness, at least consensus candidates, would have to be
kept at their habitat's temperature. This implies that any such experiment
would not be able to distinguish the linear from the non-linear dynamics,
as we would always have an irreversible entanglement with a thermal
bath. Therefore, any experiment trying to falsify CCCH on the
basis of its different dynamics is doomed. 
\begin{acknowledgements}
This is a continuation of our work with Pat Suppes, who passed away
in November 2014. This research was partially supported by the Patrick
Suppes Gift Fund for the Suppes Brain Lab. Pat's support to this paper
is gratefully acknowledged, as well as John Perry's hospitality while
JAB visited the Center for the Explanation of Consciousness at CSLI, 
Stanford University, where part of this work was conducted. We also 
thank Henry Stapp, Bas van Fraassen, and the anonymous referees 
for comments and suggestions. 
\end{acknowledgements}

\bibliographystyle{spphys}

\end{document}